\documentclass[aps,pra,showpacs,twocolumn,groupedaddress]{revtex4}
\usepackage{bm,dcolumn,amsmath,graphicx,epstopdf}

\newcommand{\eref}[1]{Eq.~(\ref{#1})}

\begin{document}

\title{Energy shift due to anisotropic blackbody radiation}
\author{V. V. Flambaum$^1$, S. G. Porsev$^{2,3}$, and M. S. Safronova$^{2,4}$}

\affiliation{$^1$School of Physics, University of New South Wales, Sydney, NSW 2052, Australia}
\affiliation{$^2$Department of Physics and Astronomy, University of Delaware, Newark, Delaware 19716, USA}
\affiliation{$^3$Petersburg Nuclear Physics Institute, Gatchina, Leningrad district 188300, Russia}
\affiliation{$^4$Joint Quantum Institute, NIST and the University of Maryland, College Park, Maryland 20742, USA}

\date{\today}

\begin{abstract}
In many applications a source of the blackbody radiation (BBR)
can be highly anisotropic. This leads to the BBR shift
that depends on tensor polarizability and  on the projection of the
total angular momentum of ions and atoms in a trap. We derived a formula for the
anisotropic BBR shift and performed numerical calculations of this effect for
Ca$^+$ and Yb$^+$  transitions of experimental interest.
These ions were used for a design of high-precision atomic clocks,
fundamental physics tests such as the search for the Lorentz invariance violation and
space-time variation of the fundamental constants, and quantum information. Anisotropic BBR shift  may be one of the major
systematic effect in these experiments.
\end{abstract}


\maketitle

\section{Introduction}
The past five years brought remarkable improvements in both accuracy and stability of atomic clocks
~\cite{NicCamHut15,BloNicWil14,UshTakDas15,GodNisJon14,HunLipTam14}.
Development of ultra-precise clocks is important for a wide range of applications, including design of absolute gravimeters and gravity
gradiometers for geophysical monitoring and research, gravity aided navigation, improved timekeeping and synchronization capabilities,
tests of fundamental physics such as Einstein’s theory of relativity, search for variation of fundamental constants through time, space,
or coupling to gravitational fields, and exploration of strongly correlated quantum many-body systems~\cite{LudBoyYe15}.

Performing fundamental tests with the atomic clocks and other precision atomic, molecular and optical (AMO) technologies leads to ever
increasing requirements for the understanding and control of the systematic errors. Moreover, a number of novel fundamental physics AMO
experiments including searches for  ultralight (sub-eV) axions, axion-like pseudoscalar and scalar dark matter~\cite{budker,stadnik1,stadnik2} and topological defect dark matter~\cite{DerPos14} have been carried out or proposed, requiring improved  understanding of systematic effects in AMO systems.

One of the major experimental and theoretical problems in improving the atomic clock accuracy is a precision determination
of atomic clock frequency shift due to blackbody radiation (BBR). In recent experimental works~\cite{UshTakDas15,NicCamHut15}
with $^{87}$Sr optical lattice clocks, the blackbody radiation  was identified as the primary source
of clock's uncertainties. A number of measurements and thorough analysis of all systematic effects
led to reduction of the Sr clock total uncertainty to the level of $2.1 \times 10^{-18}$ in fractional frequency units.
However,  65\% of the clock uncertainty budget was still due to the BBR shift \cite{NicCamHut15}.

A number of other experiments with trapped ions and atoms are sensitive to the BBR effects, including recent tests
of  local Lorentz invariance (LLI)
violation in the electron-photon sector with trapped  Ca$^+$~\cite{PruRamPor15} ions.
Theories aimed at unifying gravity with quantum physics suggest that Nature violates Lorentz symmetry at the Planck scale
while suppressing its violation at experimentally achievable energy scales \cite{KosPot95}. The minimal O(1) suppression
may lead to Lorentz-violating effects appearing beyond $10^{-17}$ sensitivity level, determined by the ratio of electroweak
and Planck scales.  Thus, high-precision experiments with atomic systems \cite{HohLeeBud13,PruRamPor15} provide an
important route to search for Lorentz violation at low energies.
The LLI experiments with trapped ions may be particulary sensitive to anisotropic BBR shift since they are based
on monitoring the energy difference between different Zeeman substates as described below. In these experiments,
anisotropic BBR shift may become a limiting factor for the ultimate accuracy of the Lorentz violation tests in the
electron-photon sector \cite{DzuFlaSaf15} and this work is strongly motivated by these fundamental studies. BBR effects may also
become a source of decoherence in larger-scale quantum information experiments with trapped ions due to
a change in the environmental temperature or temperature gradients during the computation.

Calculations of blackbody radiation shifts are usually done assuming that the BBR radiation is isotropic.

A detailed consideration of the isotropic BBR effect in conventional electric dipole approximation was carried out
in~\cite{FarWin81}. Multipolar theory of isotropic BBR shift of atomic energy levels (as well as its implications
for optical lattice clocks) was developed in~\cite{PorDer06}.
However, in practice the source of BBR can be highly anisotropic and even may have a small angular size. As a result,
a lot of experimental efforts is required  to make the BBR field uniform and with a
known temperature~\cite{NicCamHut15,UshTakDas15,BelHinPhi14}.
For this reason a proper calculation of the anisotropic BBR shift is necessary. Also, this effect can be of interest by itself as
a physical phenomenon. We note that the problem of anisotropic BBR effect, discussed in the present work, does not arise
in experiments with alkaline-earth atoms aiming to create an atomic clock in the $^1\!S_0 -\, ^3\!P^o_0$ transition because the
states with total angular momenta $J=0$ are involved.
\section{General formalism}
The isotropic BBR shift of an energy level is proportional
to scalar static polarizability of this level \cite{FarWin81,PorDer06}. In the case of anisotropic blackbody radiation,
an additional contribution that depends on the projection of a thermal photon wave vector $\boldsymbol{k}$ to the $z$ axis arises.
We show below that it is determined by the tensor polarizability of the level.
This contribution is particularly important when we consider BBR frequency shift of a $|JM \rangle$ - $|JM' \rangle$ transition,
where $M$ and $M'$ are the projections of the total angular momentum ${\bf J}$ to the $z$ axis,
 and we assume $J$ and $M$ to be good quantum numbers for the atomic states. The scalar polarizabilities
of the Zeeman substates $|JM \rangle$ and $|JM' \rangle$ are practically identical differing only due to a very small
difference in the energy denominators. As a result, the isotropic BBR frequency shift is completely negligible in
this case. In contrast, the anisotropic BBR shift of the energy level depends on $M^2$  and can be noticeably different
for the different $M$ substates. The same issue arises for the hyperfine states with different
$M_F$.

Such a systematic effect arose in a recent record-high precision experiment aimed at the  search for local Lorentz invariance
violation in the electron-photon sectors using a superposition of two Ca$^+$ ions \cite{PruRamPor15}.
In the experiment, the energy difference between the $M=1/2$ and 5/2 substates of the $3d~^2\!D_{5/2}$ multiplet,
monitored over 23 h served as a probe of Lorentz-violating effects.

The anisotropic BBR shift produces a differential shift between $M=1/2$ and 5/2 states mimicking
the Lorentz-violating effects. Thus, anisotropic BBR is an important systematic effect.
It was demonstrated in \cite{DzuFlaSaf15} that a factor of $10^5$ higher sensitivity to Lorentz violation may be
achieved with a similar experimental scheme with  Yb$^+$ by monitoring the ($4f^{13}\,6s^2$) $^2\!F_{7/2,M=7/2} -\, ^2\!F_{7/2,M=1/2}$
frequency difference. Since this experiment will probe LLI at much higher sensitivity, study of anisotropic BBR is needed as
it can be a major systematic effect for such an experiment.

We note that singly ionized ytterbium $^{171}$Yb$^+$ with ultranarrow optical $^2\!S_{1/2}$ - $^2\!D_{3/2}$ and
$^2\!S_{1/2}$ - $^2\!F_{7/2}$ transitions is also being pursued  for a realization of an optical atomic clock and search
for the temporal variation of the fine-structure constant $\alpha$ and the proton-to-electron
mass ratio $m_p/m_e$~\cite{GodNisJon14,HunLipTam14}.

The problem of anisotropic BBR shift of energy levels is practically unexplored so far, but may cause systematic effects
in a variety of experiments. In this work, we derived a general formula for the BBR shift of an energy level produced by a
point-like source. A generalization to a finite source is obtained by integration over angles of emitted thermal photons.
The result is expressed in terms of the scalar and tensor polarizabilities of the atomic level. We also performed numeric
calculation of the BBR frequency shifts for the Ca$^+$ $^2\!D_{5/2,M=5/2}$ - $^2\!D_{5/2,M=1/2}$ transition
and Yb$^+$ $^2\!F_{7/2,M=7/2}$ - $^2\!F_{7/2,M=1/2}$ transitions due to their relevance to searches for Lorentz violation.

An interaction of an atom in the state $|0\rangle$ with the electric field of a thermal photon emitted to a solid angle
$d\Omega $ leads to a blackbody radiation shift of an energy level. After integration over
photon frequency, the BBR shift of the energy level $|0\rangle$ can be written as%
\begin{equation}
\frac{dE}{d\Omega }=A \sum_{\epsilon} \sum_{i,k=1}^{3}
                      \alpha_{ik} \epsilon_{i}\epsilon _{k}^{\ast }.
\label{delE}
\end{equation}%
Here, we use a three-dimensional transverse gauge for photon polarization
$\epsilon_{\mu }=(0,\boldsymbol{\epsilon })$ with polarization $\boldsymbol{\epsilon }$ normalized to the unit.
Since photons are transverse, in this gauge $\boldsymbol{k\epsilon }=0$.
The elements of the symmetric tensor $\alpha_{ik}$ are defined as
\begin{equation}
\alpha _{ik}=2\sum_{m}\frac{\langle 0|d_{i}|m\rangle \langle m|d_{k}|0\rangle }{\omega_{m0}} ,
\label{aik}
\end{equation}%
where  $\mathbf{d}=-\mathbf{r}$ is the electric dipole moment operator and
$\omega_{m0}\equiv E_{m}-E_{0}$ is the difference between energy levels of
the intermediate and $|0\rangle$ states. We use atomic units, i.e.,  $|e|=\hbar =m_{e}=1$.
An explicit form of the factor $A$ is not
important for the following derivation and we will restore it later.

In the following we discuss only BBR effect caused by the electric field. The BBR caused by a
magnetic field was considered in Ref.~\cite{ItaLewWin82} for a number of monovalent ions and proved to
be negligible. Using the multipolar theory of blackbody radiation, developed in~\cite{PorDer06},
one can show that for the transitions in the Ca$^+$ and Yb$^+$ ions, which will be discussed below,
this effect can also be neglected.

The electric dipole static polarizability of an atom in the state $|0\rangle$ is defined as
\begin{equation}
{\alpha_{\rm pol}} \equiv \alpha_{zz} = 2 \sum_m \frac{|\langle 0 |d_z| m \rangle|^2}{\omega_{m0}}.
\label{alpha}
\end{equation}%
It can be conveniently decomposed into scalar and tensor parts:
${\alpha_{\rm pol}} = \alpha_s + \alpha_t$ with the scalar polarizability $\alpha_s$ given by
\begin{equation}
\alpha_s = \frac{1}{3} \sum_i \alpha_{ii} =
\frac{2}{3} \sum_m \frac{|\langle 0 |\mathbf{d}| m \rangle|^2}{\omega_{m0}}.
\label{alphas}
\end{equation}

The summation over photon polarizations in \eref{delE} is carried out using
\begin{equation*}
\sum_\epsilon {\epsilon_i \epsilon_k^{*}} = \delta_{ik} - n_i n_k,
\end{equation*}%
where $\boldsymbol{n}\equiv \boldsymbol{k}/k$.
 Then, Eq.~(\ref{delE}) is reduced to
\begin{eqnarray*}
\frac{dE}{d\Omega } &=& A \left[ \sum_{i}\alpha _{ii}-\sum_{i,k} \alpha_{ik} n_{i}n_{k}\right] . \\
&=& A \left[ 3\alpha _{s}-\sum_{i,k}\alpha _{ik}n_{i}n_{k} \right] .
\end{eqnarray*}%

The BBR shift in our case depends on the angle $\theta$ between the direction of the photon momentum $\boldsymbol{k}$ and
the quantization axis $z$, defined by the direction of the magnetic field.
It is convenient to choose the vector $\boldsymbol{k}$ in the $xz$ plane, i.e., $k_{y}=0$.
Taking into account Eq.~(\ref{alphas}) and noting that the product of the matrix elements
$\langle 0 |d_x| n \rangle \langle n |d_z| 0 \rangle = 0$ and, respectively, $\alpha_{zx} = \alpha_{xz}=0$, we obtain%
\begin{eqnarray*}
\frac{dE}{d\Omega } &=& A \left[ 3\alpha _{s}-\alpha _{xx}n_{x}^{2}-\alpha_{zz}n_{z}^{2}\right]  \\
&=& A \left[ 3\alpha _{s}-\alpha _{xx}\sin ^{2}\theta -\alpha _{zz}\cos^{2}\theta \right] .
\end{eqnarray*}%
Accounting for the fact that $\alpha_{xx} = \alpha_{yy}$ and, hence,
\begin{equation}
\alpha _{xx}=\frac{3\alpha _{s}-\alpha _{zz}}{2},
\label{axx}
\end{equation}%
we express $dE/d\Omega$ through $\alpha_s$, $\alpha_t$, and $\cos^{2}\!\theta $.
After simple transformations we arrive at%
\begin{equation}
\frac{dE}{d\Omega} = A' \left[ \alpha_s + \frac{1 - 3 \cos^2\!\theta}{4} \alpha_t \right] ,
\label{dE}
\end{equation}
 where $A'= 2A$. The factor $A'$ can be easily determined, if we note that after integrating
over $d\Omega = {\rm sin} \theta\, d\theta\, d\phi$ the second term in \eref{dE} disappears and we
obtain $\Delta E = 4\pi A' \alpha_s$. On the other hand, we have to
arrive at the standard formula for \textit{isotropic} BBR shift which,
neglecting dynamic corrections, is given by \cite{PorDer06}
\begin{equation*}
\Delta E = -\frac{2}{15} (\alpha \pi)^3 T^4 \alpha_s,
\end{equation*}%
where the temperature $T$ is given in a.u..
Finally, we obtain%
\begin{eqnarray}
&&dE = -\frac{2}{15} (\alpha \pi)^3 T^4 \nonumber \\
&&\times \left[ \alpha_s + \frac{1-3\cos^2\!\theta}{4}\,
\frac{3M^2 - J(J+1)}{J (2J-1)} \alpha_2 \right] \frac{d\Omega}{4\pi}.
\label{dEfin}
\end{eqnarray}%
Here we represent the tensor part $\alpha _{t}$ by%
\begin{equation}
\alpha _{t}=\frac{3M^2 - J(J+1)}{J(2J-1)}\alpha_2,
\label{al2}
\end{equation}%
where $\alpha_2$ is the tensor polarizability of the state $|0\rangle$.

It may be instructive to present a different derivation of Eq.~(\ref{dE}),
starting again from~\eref{delE}. Assuming the polarization vectors $\epsilon_{1,2}$ to be
real we can write
\begin{equation}
\frac{dE}{d\Omega }= A \sum_{i = 1}^2 \left( \alpha_{xx} \epsilon_{ix}^2
+ \alpha_{yy} \epsilon_{iy}^2 + \alpha_{zz} \epsilon_{iz}^2 \right).
\label{del1E}
\end{equation}%
Taking into account that $\alpha_{xx} = \alpha_{yy}$ and using the normalization condition 
$\epsilon_{ix}^2 + \epsilon_{iy}^2 + \epsilon_{iz}^2 =1$ and~\eref{axx}, after simple
transformations, we obtain
\begin{equation}
\frac{dE}{d\Omega }=
A \sum_{i = 1}^2 \left( \alpha_s + \frac{3\, {\rm cos}^2 \theta_i -1}{2}\alpha_t \right),
\label{del2E}
\end{equation}%
where $\theta_i$ is the angle between the photon polarization vector $\boldsymbol{\epsilon}_i$ and the $z$ axis.

Summing up over index $i$ in~\eref{del2E}, and using condition
\begin{equation}
{\rm cos}^2 \theta_1 + {\rm cos}^2 \theta_2 + {\rm cos}^2 \theta = 1,
\label{cos}
\end{equation}
which is valid because the vectors $\boldsymbol{\epsilon}_1$, $\boldsymbol{\epsilon}_2$, and
$\boldsymbol{k}$ are mutually orthogonal, we arrive at Eq.~(\ref{dE}).
\section{Anisotropic BBR shift for $^{2S+1}\!L_{J,M} - \,^{2S+1}\!L_{J,M'}$ transitions}
We now apply \eref{dEfin} to the case of a $|JM \rangle - |JM' \rangle$
transition between the ionic or atomic Zeeman sublevels. As we discussed above,
the isotropic BBR shift, proportional to the scalar part of the polarizability, is very
small for such a transition  because it results only from a small difference between $|JM \rangle$ and $|JM' \rangle$ energy levels.
The main effect comes from the tensor part of the polarizability.

Using Eqs.~(\ref{dEfin}) and (\ref{al2}) we write the $|JM \rangle - |JM' \rangle$
transition frequency BBR shift $dE_{\rm t}$ as
\begin{eqnarray}
dE_{\rm t} &\equiv& dE_{JM} - dE_{JM'} \nonumber \\
&\approx& \frac{(\alpha\pi)^3 T^4}{10}  (3\cos^2 \!\theta-1) \,
\frac{M^2-{M'}^2}{J(2J-1)}\, \alpha_2 \, \frac{d\Omega}{4\pi} .
\label{dEt}
\end{eqnarray}

Integration of \eref{dEt} over fixed solid angle $\Omega_1$ leads to the BBR shift, corresponding
to a maximal anisotropy 100\%, when the BBR is emitted to this solid angle and there is no BBR from
the remainder.

Let us consider a more realistic case when a certain portion of photons is emitted to
the solid angle $\Omega_1$ at the temperature $T_1$ and another portion of photons is emitted to
the solid angle $\Omega_2$ at the temperature $T_2$, so that $\Omega_1 + \Omega_2 = 4\pi$.
The corresponding differential BBR shifts, which we designate as $dE^{(k)}_{\rm t}$ ($k=1,2$),
are given by \eref{dEt} with $T = T_k$.

Then, the total BBR shift can be found as
\begin{eqnarray}
\Delta E_{\rm t} &=& \left[ \int_0^{\Omega_1} \frac{dE^{(1)}_{\rm t}}{d\Omega} +
                   \int_{\Omega_1}^{4\pi} \frac{dE^{(2)}_{\rm t}}{d\Omega} \right] d\Omega \nonumber \\
&=& \int_0^{\Omega_1} \left[ \frac{dE^{(1)}_{\rm t}}{d\Omega} - \frac{dE^{(2)}_{\rm t}}{d\Omega} \right] d\Omega .
\label{del_E}                   
\end{eqnarray}

Performing integration in~\eref{del_E} over azimuthal angle $\varphi$ from zero to $2\pi$
and over $\theta$ from zero to a fixed value $\theta_1$, we obtain%
\begin{eqnarray}
\Delta E_{\rm t} &\approx& \frac{(\alpha\pi)^3 (T_1^4 - T_2^4)}{20} \,\frac{M^2-{M'}^2}{J(2J-1)}\, \alpha_2 \nonumber \\
&\times& \cos \theta_1 (1-\cos^2\! \theta_1) .
\label{DeltaE}
\end{eqnarray}
As seen from \eref{DeltaE}, $\Delta E_{\rm t}$ turns to zero if $T_1 = T_2$, as it should be
because it corresponds to the isotropic BBR case.

It follows from~\eref{DeltaE} that the BBR shift is $\sim \theta_1^2$ when $\theta_1$ is small.
Thus, this shift is greatly reduced with decrease in the solid angle to which the thermal
photons are emitted. On the other hand, the anisotropic BBR shift is equal to zero when $\theta_1 = 90^\circ$.
When $\theta_1$ changes from 30 to $60^\circ$, $\cos \theta_1 (1-\cos^2\! \theta_1)$ changes from 0.22 to 0.38.

Below we consider the anisotropic BBR shifts for the $^2\!D_{5/2,M=5/2}$ - $^2\!D_{5/2,M=1/2}$ transition in Ca$^+$
and for the $^2\!F_{7/2,M=7/2}$ - $^2\!F_{7/2,M=1/2}$ transition in Yb$^+$.

\subsection{Anisotropic BBR shift for the $^2\!D_{5/2,M=5/2} -\,^2\!D_{5/2,M=1/2}$ transition in Ca$^+$.}

In a recent paper~\cite{PruRamPor15}, the $^{2}\!D_{5/2, M=5/2}$ - $^{2}\!D_{5/2, M=1/2}$ transition in Ca$^{+}$ was used to
search for Lorentz invariance violation at a level comparable to the ratio between the electroweak and Planck energy scales.

Using \eref{DeltaE}, we estimate the anisotropic BBR shift for this transition.
The most accurate value of the tensor polarizability for the $^{2}\!D_{5/2}$ state was obtained in Ref.~\cite{SafSaf11},
$\alpha_{2}=-24.51(29)$ a.u..
To illustrate dependence of the BBR shift from the temperatures $T_1$ and $T_2$, we find $\Delta E_{\rm t}$ for
three values of $T_1$ (500, 420, and 350 K) and $T_2$ = 300 K. Substituting these values in \eref{DeltaE},
taking into account that for the room temperature 300~K $(\alpha\pi)^3 T^4/20 \approx 4.9069 \times 10^{-19}$ a.u.,
and expressing final results in Hz, we obtain 
\begin{eqnarray}
\Delta E_{\rm t} &\approx& \cos \theta_1 (1-\cos^2\! \theta_1) \nonumber \\
&\times& \left\lbrace
\begin{array}{l}
-0.319\, \text{(Hz)}, \, T_1 = 500\, {\rm K},  \\
-0.135\, \text{(Hz)}, \, T_1 = 420\, {\rm K},  \\
-0.040\, \text{(Hz)}, \, T_1 = 350\, {\rm K}.  
\end{array}
\right.
\label{dE_Ca}
\end{eqnarray}
\begin{figure} [!htb]
\includegraphics[width=\linewidth]{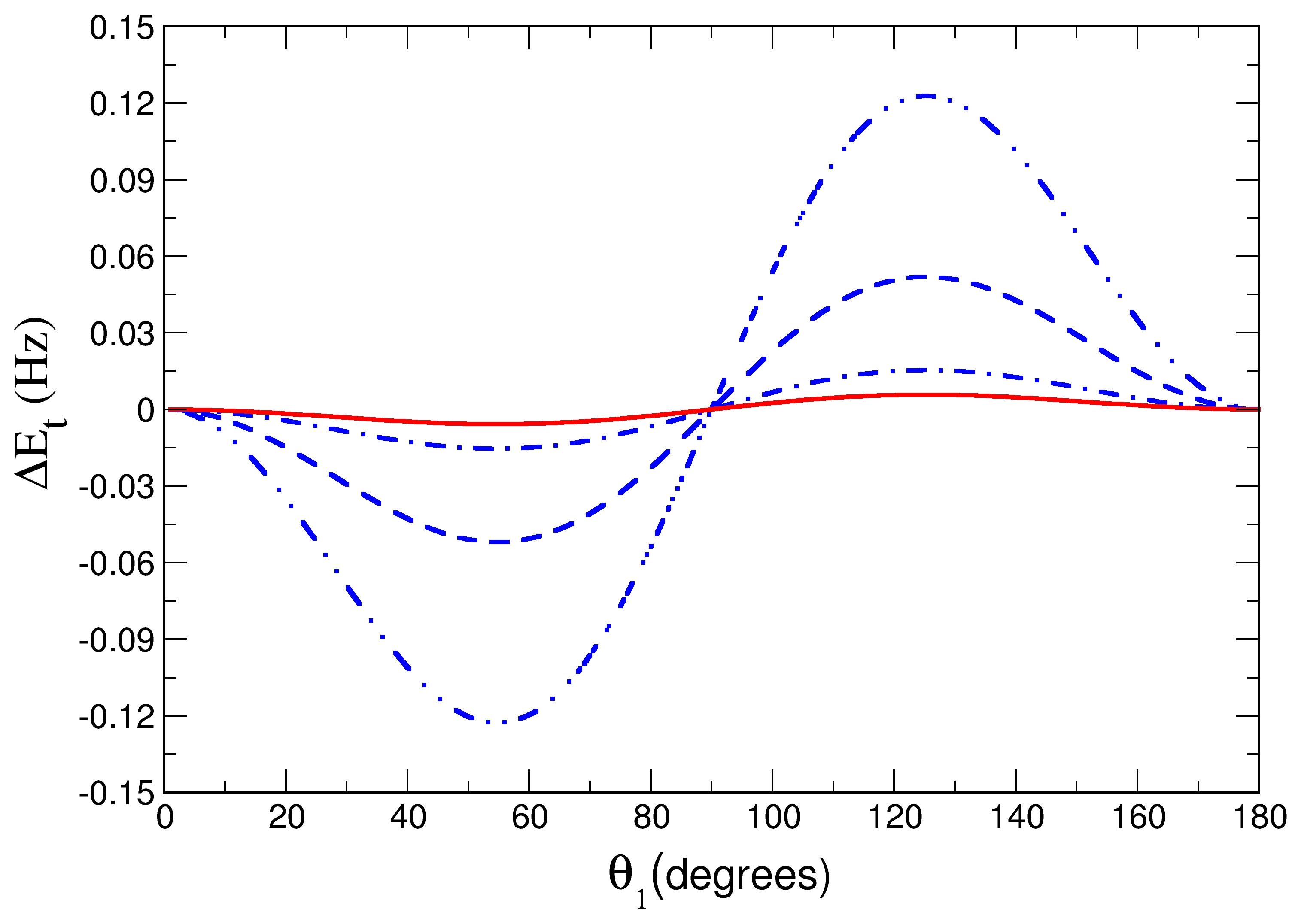}
\caption{(Color online) Dependence of $\Delta E_{\rm t}$ from the angle $\theta_1$
for Ca$^+$ (see Eq.~(\ref{dE_Ca})) is represented by (blue) dash-dot-dotted line for $T_1$ = 500 K,
by dashed line for $T_1$ = 420 K, and by dot-dashed line for $T_1$ = 350 K.
Dependence of $\Delta E_{\rm t}$ from $\theta_1$ for Yb$^+$ (see Eq.~(\ref{dE_Yb})) is represented by
(red) solid line for $T_1$ = 450 K. The temperature $T_2$ = 300 K in all cases.}
\label{fig:BBR}
\end{figure}

These dependences of $\Delta E_{\rm t}$ from the angle $\theta_1$ in Ca$^+$
$^{2}\!D_{5/2, M=5/2}$ - $^{2}\!D_{5/2, M=1/2}$ transition are illustrated in Fig.~\ref{fig:BBR}
by three (blue) lines. As expected, $\Delta E_{\rm t}$ is equal to zero at $\theta_1 = 0$ and $180^\circ$.
The angle $\theta_1 = 180^\circ$ corresponds to isotropic radiation. $\Delta E_{\rm t}$ also crosses zero
when $\theta_1$ is $90^\circ$. This case corresponds to isotropic radiation in the upper hemisphere.
The energy shifts due to the BBR effect are largest for $\theta_1 \approx 55^\circ$.

\subsection{Anisotropic BBR shift for the $^2\!F_{7/2,M=7/2} -\,^2\!F_{7/2,M=1/2}$ transition in Yb$^+$.}

Recent work ~\cite{DzuFlaSaf15} identified several factors affecting the precision of the local Lorentz invariance
tests with trapped ions. The two most important factors are the lifetime of the excited atomic state used in the Lorentz
invariance probe and sensitivity of this state to the Lorentz invariance violation effect, i.e., the size of the matrix element
of the corresponding operator. Both features are supplied by the metastable $4f^{13} 6s^2 \,^2\!F_{7/2}$ state of the Yb$^+$ ion, and
the $^{2}\!F_{7/2,M=7/2}$ - $^{2}\!F_{7/2,M=1/2}$ transition  was proposed as the probe of Lorentz-violating effects.
To estimate this BBR shift we need to evaluate the value of the tensor polarizability $\alpha_2 (^2\!F_{7/2})$. 
We carried out calculations in the framework of 15-electron configuration interaction (CI) method,
following the approach described in Ref.~\cite{PorSafKoz12Ybp}.

The main features of this approach are briefly described below.
All electrons are divided into the core and valence electrons. In our case [$1s^2$,...,$5p^6$] are the core electrons
while 15 outer electrons belong to the valence subspace.

In the framework of the CI method we solve the eigenvalue problem
\begin{equation}
H_{\rm CI} \Phi = E_{\rm CI} \Phi,
\label{Sch}
\end{equation}
where the many-electron wave functions $\Phi$ belong to the valence subspace and are presented as a linear
combination of Slater determinants,
\begin{equation}\label{Phi1}
\Phi = \sum_{{\rm det}_I} C_I | {\rm det}_I \rangle.
\end{equation}

The CI Hamiltonian can be written as
\begin{eqnarray}
H_{\rm CI} = E_{\rm core} + \sum_{i>N_{\rm core}}{h^{\rm CI}_i} +
\sum_{j>i>N_{\rm core}} V_{ij}\,,
\label{e2}
\end{eqnarray}
where $N_{\rm core}$ is the number of core electrons,
$E_{\rm core}$ is the energy of the core which includes kinetic
energy of the core electrons, Coulomb energy of their interaction
with the nucleus and potential energy of the core-core electrostatic
interaction. The core-valence interaction, kinetic energy of the valence electrons
and their interaction with the nucleus are included in the one-electron operators $h^{\rm CI}_i$.
The last term in Eq.~(\ref{e2}) accounts for the interaction between
valence electrons.

We start from solving the Dirac-Fock equations for the [$1s^2$,...,$4f^{13} 6s^2$] configuration. Then the $6p_j$ and $5d_j$ orbitals are
constructed for the $4f^{13} 6s 6p$ and $4f^{13} 6p5d$ configurations, correspondingly. The basis set used in the CI calculations included
also virtual orbitals up to $8s$, $8p$, $7d$, $7f$, and $5g$.
We form configuration space by allowing single and double excitations for the odd-parity states from the $4f^{14} 6p$, $4f^{13} 6s^2$ and
$4f^{13} 5d6s$ configurations and for the even-parity states from the $4f^{14} 6s$, $4f^{13} 6s6p$ and
$4f^{13} 6p5d$ configurations to the orbitals of the basis set listed above.

 Solving the relativistic multiparticle Schr\"{o}dinger equation,~\eref{Sch} gives the eigenvector of the $4f^{13} 6s^2$~$^2\!F_{7/2}$ state which we use to determine the tensor polarizability of this state.

Using formalism of the reduced matrix elements we can write the
expression for the tensor polarizability of the state $\Phi_0$ with total
angular momentum $J$ as~\cite{KozPor99}
\begin{eqnarray}
&&\alpha_2 = 4 \left(\frac{5J(2J-1)}{6(2J+3)(2J+1)(J+1)}\right)^{1/2} \nonumber \\
&\times& \sum_{n} (-1)^{J+J_n}
\left\{
\begin{array}{ccc} J & 1 & J_n \\ 1 & J & 2
\end{array}
\right\}  
\frac{|\langle \Phi_0||d||\Phi_n\rangle|^2} {E_n-E_a},
\label{at}
\end{eqnarray}
where $J_n$ is the total angular momentum of the intermediate state $\Phi_n$.

A direct summation over all intermediate states in~\eref{at} requires a knowledge of the complete set of
eigenstates of the Hamiltonian~(\ref{Sch}). Practically, this is impossible when dimension of a CI space exceeds
few thousand determinants, as in our case. To find the electric dipole tensor polarizability of the $4f^{13} 6s^2$~$^2\!F_{7/2}$ state
we use the method of solution of an inhomogeneous equation, described in detail in~\cite{KozPor99}. The random-phase-approximation
corrections are also included.

The result of our computation of the tensor polarizability is $\alpha_2 (^2\!F_{7/2}) \approx -2$ a.u..
Using this value and $T_1$ = 450 K and $T_2$ = 300 K, we obtain from \eref{DeltaE}
\begin{eqnarray}
\Delta E_{\rm t} &\approx& -0.015\, \cos \theta_1 (1-\cos^2\! \theta_1) \,\, {\rm Hz}.
\label{dE_Yb}
\end{eqnarray}

The dependence of $\Delta E_{\rm t}$ on the angle $\theta_1$ is shown in Fig.~\ref{fig:BBR}
by a (red) solid line. While the general behavior of $\Delta E_{\rm t}$ for Yb$^+$ is the same at
$\theta_1$ = 0, 90, and $180^\circ$ as in Ca$^+$, the $\Delta E_{\rm t}$ is much smaller in the Yb$^+$
transition considered here than in the Ca$^+$ one. Suppression of the anisotropic blackbody radiation in
the $^{2}\!F_{7/2,M=7/2}$ - $^{2}\!F_{7/2,M=1/2}$ Yb$^+$
transition is due to compactness of the Yb$^+$ $4f$ orbital, resulting in
the value of tensor polarizability, which is  an order of magnitude smaller than that for the $^2\!D_{5/2}$
state of Ca$^+$. Therefore, the anisotropic BBR shift  is strongly suppressed for  transition between substates of
the Yb$^+$ $^2F_{7/2}$ multiplet, and for $T_1$ = 450 and $T_2$ = 300 K its maximal (absolute) value at
$\theta_1 \approx 55^\circ$ is equal to 5.8 mHz.

To conclude, we derived the formula for the
anisotropic BBR shift of an energy level and performed numerical calculations of this effect for Ca$^+$ and Yb$^+$
transitions of interest for study of Lorentz violation. We demonstrated  that this effect strongly depends on the
magnitude of the tensor polarizability of the level. In high-precision experiments, the anisotropic BBR
can be a major systematic effect that should be specifically addressed in determining experimental uncertainties.\\

\noindent M.S.S. thanks the School of Physics at University of New South Wales (UNSW), Sydney, Australia  for hospitality and
acknowledges support from the Gordon Godfrey Fellowship program, UNSW.
This work was supported by NSF Grants No. PHY-1404156, No. PHY-1212442, and No. PHY-1520993 and the Australian Research Council.


\begin{thebibliography}{21}
\expandafter\ifx\csname natexlab\endcsname\relax\def\natexlab#1{#1}\fi
\expandafter\ifx\csname bibnamefont\endcsname\relax
  \def\bibnamefont#1{#1}\fi
\expandafter\ifx\csname bibfnamefont\endcsname\relax
  \def\bibfnamefont#1{#1}\fi
\expandafter\ifx\csname citenamefont\endcsname\relax
  \def\citenamefont#1{#1}\fi
\expandafter\ifx\csname url\endcsname\relax
  \def\url#1{\texttt{#1}}\fi
\expandafter\ifx\csname urlprefix\endcsname\relax\def\urlprefix{URL }\fi
\providecommand{\bibinfo}[2]{#2}
\providecommand{\eprint}[2][]{\url{#2}}

\bibitem[{\citenamefont{Nicholson et~al.}(2015)\citenamefont{Nicholson,
  Campbell, Hutson, Marti, Bloom, McNally, Zhang, Barrett, Safronova, Strouse
  et~al.}}]{NicCamHut15}
\bibinfo{author}{\bibfnamefont{T.~L.} \bibnamefont{Nicholson}},
  \bibinfo{author}{\bibfnamefont{S.~L.} \bibnamefont{Campbell}},
  \bibinfo{author}{\bibfnamefont{R.~B.} \bibnamefont{Hutson}},
  \bibinfo{author}{\bibfnamefont{G.~E.} \bibnamefont{Marti}},
  \bibinfo{author}{\bibfnamefont{B.~J.} \bibnamefont{Bloom}},
  \bibinfo{author}{\bibfnamefont{R.~L.} \bibnamefont{McNally}},
  \bibinfo{author}{\bibfnamefont{W.}~\bibnamefont{Zhang}},
  \bibinfo{author}{\bibfnamefont{M.~D.} \bibnamefont{Barrett}},
  \bibinfo{author}{\bibfnamefont{M.~S.} \bibnamefont{Safronova}},
  \bibinfo{author}{\bibfnamefont{G.~F.} \bibnamefont{Strouse}},
  \bibnamefont{et~al.}, \bibinfo{journal}{Nature Comm.}
  \textbf{\bibinfo{volume}{6}}, \bibinfo{pages}{6896} (\bibinfo{year}{2015}).

\bibitem[{\citenamefont{Bloom et~al.}(2014)\citenamefont{Bloom, Nicholson,
  Williams, Campbell, Bishof, Zhang, Zhang, Bromley, and Ye}}]{BloNicWil14}
\bibinfo{author}{\bibfnamefont{B.~J.} \bibnamefont{Bloom}},
  \bibinfo{author}{\bibfnamefont{T.~L.} \bibnamefont{Nicholson}},
  \bibinfo{author}{\bibfnamefont{J.~R.} \bibnamefont{Williams}},
  \bibinfo{author}{\bibfnamefont{S.~L.} \bibnamefont{Campbell}},
  \bibinfo{author}{\bibfnamefont{M.}~\bibnamefont{Bishof}},
  \bibinfo{author}{\bibfnamefont{X.}~\bibnamefont{Zhang}},
  \bibinfo{author}{\bibfnamefont{W.}~\bibnamefont{Zhang}},
  \bibinfo{author}{\bibfnamefont{S.~L.} \bibnamefont{Bromley}},
  \bibnamefont{and} \bibinfo{author}{\bibfnamefont{J.}~\bibnamefont{Ye}},
  \bibinfo{journal}{Nature} \textbf{\bibinfo{volume}{506}}, \bibinfo{pages}{71}
  (\bibinfo{year}{2014}).

\bibitem[{\citenamefont{Ushijima et~al.}(2015)\citenamefont{Ushijima, Takamoto,
  Das, Ohkubo, and Katori}}]{UshTakDas15}
\bibinfo{author}{\bibfnamefont{I.}~\bibnamefont{Ushijima}},
  \bibinfo{author}{\bibfnamefont{M.}~\bibnamefont{Takamoto}},
  \bibinfo{author}{\bibfnamefont{M.}~\bibnamefont{Das}},
  \bibinfo{author}{\bibfnamefont{T.}~\bibnamefont{Ohkubo}}, \bibnamefont{and}
  \bibinfo{author}{\bibfnamefont{H.}~\bibnamefont{Katori}},
  \bibinfo{journal}{Nat. Photonics} \textbf{\bibinfo{volume}{9}},
  \bibinfo{pages}{185} (\bibinfo{year}{2015}).

\bibitem[{\citenamefont{Godun et~al.}(2014)\citenamefont{Godun, Nisbet-Jones,
  Jones, King, Johnson, Margolis, Szymaniec, Lea, Bongs, and
  Gill}}]{GodNisJon14}
\bibinfo{author}{\bibfnamefont{R.~M.} \bibnamefont{Godun}},
  \bibinfo{author}{\bibfnamefont{P.~B.~R.} \bibnamefont{Nisbet-Jones}},
  \bibinfo{author}{\bibfnamefont{J.~M.} \bibnamefont{Jones}},
  \bibinfo{author}{\bibfnamefont{S.~A.} \bibnamefont{King}},
  \bibinfo{author}{\bibfnamefont{L.~A.~M.} \bibnamefont{Johnson}},
  \bibinfo{author}{\bibfnamefont{H.~S.} \bibnamefont{Margolis}},
  \bibinfo{author}{\bibfnamefont{K.}~\bibnamefont{Szymaniec}},
  \bibinfo{author}{\bibfnamefont{S.~N.} \bibnamefont{Lea}},
  \bibinfo{author}{\bibfnamefont{K.}~\bibnamefont{Bongs}}, \bibnamefont{and}
  \bibinfo{author}{\bibfnamefont{P.}~\bibnamefont{Gill}},
  \bibinfo{journal}{Phys. Rev. Lett.} \textbf{\bibinfo{volume}{113}},
  \bibinfo{pages}{210801} (\bibinfo{year}{2014}).

\bibitem[{\citenamefont{Huntemann et~al.}(2014)\citenamefont{Huntemann,
  Lipphardt, Tamm, Gerginov, Weyers, and Peik}}]{HunLipTam14}
\bibinfo{author}{\bibfnamefont{N.}~\bibnamefont{Huntemann}},
  \bibinfo{author}{\bibfnamefont{B.}~\bibnamefont{Lipphardt}},
  \bibinfo{author}{\bibfnamefont{C.}~\bibnamefont{Tamm}},
  \bibinfo{author}{\bibfnamefont{V.}~\bibnamefont{Gerginov}},
  \bibinfo{author}{\bibfnamefont{S.}~\bibnamefont{Weyers}}, \bibnamefont{and}
  \bibinfo{author}{\bibfnamefont{E.}~\bibnamefont{Peik}},
  \bibinfo{journal}{Phys. Rev. Lett.} \textbf{\bibinfo{volume}{113}},
  \bibinfo{pages}{210802} (\bibinfo{year}{2014}).

\bibitem[{\citenamefont{{Ludlow} et~al.}(2015)\citenamefont{{Ludlow}, {Boyd},
  {Ye}, {Peik}, and {Schmidt}}}]{LudBoyYe15}
\bibinfo{author}{\bibfnamefont{A.~D.} \bibnamefont{{Ludlow}}},
  \bibinfo{author}{\bibfnamefont{M.~M.} \bibnamefont{{Boyd}}},
  \bibinfo{author}{\bibfnamefont{J.}~\bibnamefont{{Ye}}},
  \bibinfo{author}{\bibfnamefont{E.}~\bibnamefont{{Peik}}}, \bibnamefont{and}
  \bibinfo{author}{\bibfnamefont{P.~O.} \bibnamefont{{Schmidt}}},
  \bibinfo{journal}{Rev. Mod. Phys.} \textbf{\bibinfo{volume}{87}},
  \bibinfo{pages}{637} (\bibinfo{year}{2015}).

\bibitem[{\citenamefont{{\rm Van Tilburg} et~al.}(2015)\citenamefont{{\rm Van
  Tilburg}, Leefer, Bougas, and Budker}}]{budker}
\bibinfo{author}{\bibfnamefont{K.}~\bibnamefont{{\rm Van Tilburg}}},
  \bibinfo{author}{\bibfnamefont{N.}~\bibnamefont{Leefer}},
  \bibinfo{author}{\bibfnamefont{L.}~\bibnamefont{Bougas}}, \bibnamefont{and}
  \bibinfo{author}{\bibfnamefont{D.}~\bibnamefont{Budker}},
  \bibinfo{journal}{Phys. Rev. Lett.} \textbf{\bibinfo{volume}{115}},
  \bibinfo{pages}{011802} (\bibinfo{year}{2015}).

\bibitem[{\citenamefont{Stadnik and Flambaum}(2015)}]{stadnik1}
\bibinfo{author}{\bibfnamefont{Y.~V.} \bibnamefont{Stadnik}} \bibnamefont{and}
  \bibinfo{author}{\bibfnamefont{V.~V.} \bibnamefont{Flambaum}},
  \bibinfo{journal}{Phys. Rev. Lett.} \textbf{\bibinfo{volume}{115}},
  \bibinfo{pages}{201301} (\bibinfo{year}{2015}).

\bibitem[{sta()}]{stadnik2}
\bibinfo{note}{Y. V. Stadnik and V. V. Flambaum, arXiv:1504.01798}.

\bibitem[{\citenamefont{{Derevianko} and {Pospelov}}(2014)}]{DerPos14}
\bibinfo{author}{\bibfnamefont{A.}~\bibnamefont{{Derevianko}}}
  \bibnamefont{and}
  \bibinfo{author}{\bibfnamefont{M.}~\bibnamefont{{Pospelov}}},
  \bibinfo{journal}{Nature Physics} \textbf{\bibinfo{volume}{10}},
  \bibinfo{pages}{933} (\bibinfo{year}{2014}).

\bibitem[{\citenamefont{Pruttivarasin et~al.}(2015)\citenamefont{Pruttivarasin,
  Ramm, Porsev, Tupitsyn, Safronova, Hohensee, and H\"affner}}]{PruRamPor15}
\bibinfo{author}{\bibfnamefont{T.}~\bibnamefont{Pruttivarasin}},
  \bibinfo{author}{\bibfnamefont{M.}~\bibnamefont{Ramm}},
  \bibinfo{author}{\bibfnamefont{S.~G.} \bibnamefont{Porsev}},
  \bibinfo{author}{\bibfnamefont{I.}~\bibnamefont{Tupitsyn}},
  \bibinfo{author}{\bibfnamefont{M.~S.} \bibnamefont{Safronova}},
  \bibinfo{author}{\bibfnamefont{M.~A.} \bibnamefont{Hohensee}},
  \bibnamefont{and}
  \bibinfo{author}{\bibfnamefont{H.}~\bibnamefont{H\"affner}},
  \bibinfo{journal}{Nature} \textbf{\bibinfo{volume}{517}},
  \bibinfo{pages}{592} (\bibinfo{year}{2015}).

\bibitem[{\citenamefont{{Kosteleck{\'y}} and {Potting}}(1995)}]{KosPot95}
\bibinfo{author}{\bibfnamefont{V.~A.} \bibnamefont{{Kosteleck{\'y}}}}
  \bibnamefont{and}
  \bibinfo{author}{\bibfnamefont{R.}~\bibnamefont{{Potting}}},
  \bibinfo{journal}{\prd} \textbf{\bibinfo{volume}{51}}, \bibinfo{pages}{3923}
  (\bibinfo{year}{1995}).

\bibitem[{\citenamefont{{Hohensee} et~al.}(2013)\citenamefont{{Hohensee},
  {Leefer}, {Budker}, {Harabati}, {Dzuba}, and {Flambaum}}}]{HohLeeBud13}
\bibinfo{author}{\bibfnamefont{M.~A.} \bibnamefont{{Hohensee}}},
  \bibinfo{author}{\bibfnamefont{N.}~\bibnamefont{{Leefer}}},
  \bibinfo{author}{\bibfnamefont{D.}~\bibnamefont{{Budker}}},
  \bibinfo{author}{\bibfnamefont{C.}~\bibnamefont{{Harabati}}},
  \bibinfo{author}{\bibfnamefont{V.~A.} \bibnamefont{{Dzuba}}},
  \bibnamefont{and} \bibinfo{author}{\bibfnamefont{V.~V.}
  \bibnamefont{{Flambaum}}}, \bibinfo{journal}{Phys. Rev. Lett.}
  \textbf{\bibinfo{volume}{111}}, \bibinfo{eid}{050401} (\bibinfo{year}{2013}).

\bibitem[{Dzu()}]{DzuFlaSaf15}
\bibinfo{note}{V. A. Dzuba, V. V. Flambaum, M. S. Safronova, S. G. Porsev, T.
  Pruttivarasin, M. A. Hohensee, and H. H\"affner, Nat. Phys. (2016), doi:
  10.1038/nphys3610; arXiv:1507.06048}.

\bibitem[{\citenamefont{Farley and Wing}(1981)}]{FarWin81}
\bibinfo{author}{\bibfnamefont{J.~W.} \bibnamefont{Farley}} \bibnamefont{and}
  \bibinfo{author}{\bibfnamefont{W.~H.} \bibnamefont{Wing}},
  \bibinfo{journal}{Phys. Rev. A} \textbf{\bibinfo{volume}{23}},
  \bibinfo{pages}{2397} (\bibinfo{year}{1981}).

\bibitem[{\citenamefont{Porsev and Derevianko}(2006)}]{PorDer06}
\bibinfo{author}{\bibfnamefont{S.~G.} \bibnamefont{Porsev}} \bibnamefont{and}
  \bibinfo{author}{\bibfnamefont{A.}~\bibnamefont{Derevianko}},
  \bibinfo{journal}{Phys. Rev. A} \textbf{\bibinfo{volume}{74}},
  \bibinfo{pages}{020502(R)} (\bibinfo{year}{2006}).

\bibitem[{\citenamefont{Beloy et~al.}(2014)\citenamefont{Beloy, Hinkley,
  Phillips, Sherman, Schioppo, Lehman, Feldman, Hanssen, Oates, and
  Ludlow}}]{BelHinPhi14}
\bibinfo{author}{\bibfnamefont{K.}~\bibnamefont{Beloy}},
  \bibinfo{author}{\bibfnamefont{N.}~\bibnamefont{Hinkley}},
  \bibinfo{author}{\bibfnamefont{N.~B.} \bibnamefont{Phillips}},
  \bibinfo{author}{\bibfnamefont{J.~A.} \bibnamefont{Sherman}},
  \bibinfo{author}{\bibfnamefont{M.}~\bibnamefont{Schioppo}},
  \bibinfo{author}{\bibfnamefont{J.}~\bibnamefont{Lehman}},
  \bibinfo{author}{\bibfnamefont{A.}~\bibnamefont{Feldman}},
  \bibinfo{author}{\bibfnamefont{L.~M.} \bibnamefont{Hanssen}},
  \bibinfo{author}{\bibfnamefont{C.~W.} \bibnamefont{Oates}}, \bibnamefont{and}
  \bibinfo{author}{\bibfnamefont{A.~D.} \bibnamefont{Ludlow}},
  \bibinfo{journal}{Phys. Rev. Lett.} \textbf{\bibinfo{volume}{113}},
  \bibinfo{pages}{260801} (\bibinfo{year}{2014}).

\bibitem[{\citenamefont{Itano et~al.}(1982)\citenamefont{Itano, Lewis, and
  Wineland}}]{ItaLewWin82}
\bibinfo{author}{\bibfnamefont{W.~M.} \bibnamefont{Itano}},
  \bibinfo{author}{\bibfnamefont{L.~L.} \bibnamefont{Lewis}}, \bibnamefont{and}
  \bibinfo{author}{\bibfnamefont{D.~J.} \bibnamefont{Wineland}},
  \bibinfo{journal}{Phys. Rev. A} \textbf{\bibinfo{volume}{25}},
  \bibinfo{pages}{1233} (\bibinfo{year}{1982}).

\bibitem[{\citenamefont{Safronova and Safronova}(2011)}]{SafSaf11}
\bibinfo{author}{\bibfnamefont{M.~S.} \bibnamefont{Safronova}}
  \bibnamefont{and} \bibinfo{author}{\bibfnamefont{U.~I.}
  \bibnamefont{Safronova}}, \bibinfo{journal}{Phys. Rev. A}
  \textbf{\bibinfo{volume}{83}}, \bibinfo{pages}{012503}
  (\bibinfo{year}{2011}).

\bibitem[{\citenamefont{Porsev et~al.}(2012)\citenamefont{Porsev, Safronova,
  and Kozlov}}]{PorSafKoz12Ybp}
\bibinfo{author}{\bibfnamefont{S.~G.} \bibnamefont{Porsev}},
  \bibinfo{author}{\bibfnamefont{M.~S.} \bibnamefont{Safronova}},
  \bibnamefont{and} \bibinfo{author}{\bibfnamefont{M.~G.}
  \bibnamefont{Kozlov}}, \bibinfo{journal}{Phys. Rev. A}
  \textbf{\bibinfo{volume}{86}}, \bibinfo{pages}{022504}
  (\bibinfo{year}{2012}).

\bibitem[{\citenamefont{Kozlov and Porsev}(1999)}]{KozPor99}
\bibinfo{author}{\bibfnamefont{M.~G.} \bibnamefont{Kozlov}} \bibnamefont{and}
  \bibinfo{author}{\bibfnamefont{S.~G.} \bibnamefont{Porsev}},
  \bibinfo{journal}{Eur. Phys. J. D} \textbf{\bibinfo{volume}{5}},
  \bibinfo{pages}{59} (\bibinfo{year}{1999}).

\end{thebibliography}

\end{document}